\documentclass[aps,prl,twocolumn,showpacs,preprintnumbers,amsmath,amssymb]{revtex4}
\usepackage{amsmath,amsfonts,amssymb,latexsym}
\usepackage{epsfig}

\begin{document}
\title{Thermodynamic forces, flows, and Onsager coefficients in complex networks}
\author{Agata Fronczak, Piotr Fronczak and Janusz A. Ho\l yst}
\affiliation{Faculty of Physics and Center of Excellence for
Complex Systems Research, Warsaw University of Technology,
Koszykowa 75, PL-00-662 Warsaw, Poland}
\date{\today}

\begin{abstract}
We present Onsager formalism applied to random networks with
arbitrary degree distribution. Using the well-known methods of
non-equilibrium thermodynamics we identify thermodynamic forces
and their conjugated flows induced in networks as a result of
single node degree perturbation. The forces and the flows can be
understood as a response of the system to events, such as random
removal of nodes or intentional attacks on them. Finally, we show
that cross effects (such as thermodiffusion, or thermoelectric
phenomena), in which one force may not only give rise to its own
corresponding flow, but to many other flows, can be observed also
in complex networks.
\end{abstract}  \maketitle

Onsager relations \cite{OnsagerPR} constitute one of the most
prominent results of the traditional non-equilibrium statistical
physics \cite{Mazur,Kondepundi}. In short, they explain why and
how small perturbations of some system parameters can induce
fluctuations of other parameters.

The relations are derived from the assumption that the response of
the system, which is close to equilibrium, to small external
perturbation is the same as its response to a spontaneous
fluctuation. Since the considered systems are close to equilibrium
the change in entropy $dS$ is mainly due to entropy production
$d_iS$, the rate of which can be written as
\begin{equation}\label{RateS1}
\sigma=\frac{d_iS}{dt}=\sum_jF_jJ_j,
\end{equation}
where $F_j$ are thermodynamic forces, such as the gradient of
$1/T$, and $J_j$ are flows, such as the heat flow. In the vicinity
of thermodynamic equilibrium, the following linear relation
between the flows and the forces holds
\begin{equation}\label{FlowGen}
J_j=\sum_{i}L_{ji}F_i,
\end{equation}
where $L_{ji}$ represent the so-called phenomenological
coefficients, which have been proved to fulfil the Onsager
reciprocal relations
\begin{equation}\label{OnsagerRel}
L_{ji}=L_{ij}.
\end{equation}

Please note that the relation (\ref{FlowGen}) implies that not
only can a force such as the gradient of $1/T$ cause the heat flow
but it can also drive other flows, such as a flow of matter or an
electrical current. In other words, an entropic force $F_i$ may
not only give rise to its corresponding flux $J_i$, but to many
other fluxes $J_j$ in a dazzling variety. Moreover, due to
(\ref{OnsagerRel}), one flow $J_j$ causes the other $J_i$ in
exactly the same way and to exactly the same extent. The
thermoelectric effect is one such a cross effect. Thermodiffusion
is another example. The proliferation of fluxes described above is
the main reason why it is so difficult to perceive causality in
complex systems, in which relationships between constituents may
give rise to very complicated behaviors. Notwithstanding these
difficulties, in the paper we examine effects of the Onsager
causality in complex networks, which during the last decade have
broadened the purview of physics.

In a nutshell, real-world networks and their theoretical models
are called complex by a virtue of a set of non-trivial topological
features among which the most prominent are: heavy-tail in the
degree distribution, tendency of nodes to form clusters, small
world effect, assortativity or disassortativity among vertices,
community structure at many scales, and evidence of a hierarchical
structure (for an extensive review see Refs.~\cite{net1,net2}).
Since Onsager relations operate when the considered systems are
close to equilibrium, in the following we will concentrate on
equilibrium networks, precisely on exponential random graphs, also
known as $p*$ models, neglecting a huge class of evolving
non-equilibrium networks.

Exponential random graphs are ensemble models. They are already
well-known for mathematicians \cite{pmodels1,pmodels2}, and
recently have also aroused interest among physicists
\cite{Park04,FronczakFluct,Garla06}. As a matter of fact the
methodology behind the models directly follows the methodology
behind maximum entropy school of thermodynamics \cite{Jaynes57}.
In order to correctly define an ensemble of networks, one has to
specify a set of graphs $\mathcal{G}$ that one wants to study. In
the following we restrict ourselves to labelled simple graphs with
a fixed number of nodes $N$. Next, since the set $\mathcal{G}$ of
possible networks has been established, one has to decide what
kind of constraints should be imposed on the ensemble. The choice
may be, for example, encouraged by properties of real networks
such as high clustering, significant modularity, or scale-free
degree distribution $P(k)\sim k^{-\gamma}$. Then, one specifies
probability distribution $P(G)$ ($G\in\mathcal{G}$) over the
ensemble, which consists in maximization of the Shannon entropy
$S=-\sum_{G} P(G)\ln(G)$ subject to the given constraints. The
procedure leads to the Boltzmann-like probability distribution
\begin{equation}\label{PG}
P(G)=\frac{e^{-H(G)}}{Z},
\end{equation}
where $Z$ stands for the partition function, whereas
$H(G)=\sum_{j}\theta_jm_j(G)$ is called the graph Hamiltonian. The
set $\{m_j\}$ represents ensemble free parameters (like energy $E$
in the canonical ensemble) upon which the relevant constraints
act, and $\{\theta_j\}$ is a set of fields conjugated to these
parameters (like $\beta=(kT)^{-1}$ representing field conjugated
to the energy $E$). Further in the paper, we will consider network
ensembles characterized by a desired degree sequence
$\{h_1,h_2,\dots,h_N\}$, i.e. by the Hamiltonian of the form
\cite{Park04}
\begin{equation}\label{H1}
H(G)=\sum_{i=1}^N\theta_ik_i(G).
\end{equation}
The ensembles are formally equivalent to uncorrelated networks
with a given node degree distribution $P(k)$ \cite{FronczakFluct},
which have been repeatedly used in recent years as the simplest
(but not yet trivial!) models of real networks
\cite{New01,Pastor01,HavlinPRLe}. The Onsager formalism applied to
this ensemble will allow us to study dynamical response of the
considered networks to external perturbations.

In the following, we will study the simplest kind of perturbation
consisting in a sudden change of single node's connectivity, e.g.
$k_i(t_0)=0$. The perturbation is particularly well suited for the
Hamiltonian (\ref{H1}) because nodes degrees are ensemble free
parameters in the case. Let us also stress that the perturbation
directly corresponds to frequently discussed problems of random or
intentional removal of sites and links in complex networks, which
have been considered in relation with such important issues as:
resilience of real networks to random breakdowns, their
susceptibility to intentional attacks, and finally the issue of
cascading failures in these networks. Although, however, a number
of analysis in the field has been performed, most of them may be
classified into one of the two categories: the first one focusing
on static, percolation properties of new networks arising as a
result of a given perturbation \cite{HavlinPRLe,BANatur}, and the
second one encompassing a variety of processes which excel at
imitating specific phenomena (like clogging in the Internet) and
give some insight into dynamical behavior of the considered
networks after such a perturbation
\cite{MotterPRL,BianconiPRE,FronczakSOC}. The approach presented
in this paper does not fall into neither category. Although in the
paper we concentrate on a similar kind of perturbation the true
challenge of our approach is to present how the most fundamental
results of non-equilibrium thermodynamics can help in
understanding of complex networks. The approach is all the more
important, since it can be applied to any ensemble of networks
with an arbitrary graph Hamiltonian (\ref{PG}).

Thus, let us apply the Onsager formalism to ensemble of networks
described by the Hamiltonian (\ref{H1}). Our first aim is to
determine thermodynamic flows and forces (\ref{RateS1}) which
appear in the networks after the perturbation consisting in a
sudden change of a single node's degree. In order to do it one has
to expand the ensemble entropy $S(k_1,k_2,\dots,k_N)$ about
equilibrium as a power series in its independent variables
\begin{equation}
d_iS=S-S_{eq}=\frac{1}{2}\sum_{i,j}\frac{\partial^2 S}{\partial
k_i\partial k_j}(k_i-h_i)(k_j-h_j),
\end{equation}
where $\partial S/\partial k_i=0$. Next, computing the time
derivative of the above expression one obtains a new microscopic
expression for the rate of the entropy production
\begin{equation}\label{RateS2}
\sigma=\frac{d_iS}{dt}=-\sum_{i,j}g_{ij}(k_i-h_i)
\frac{d(k_j-h_j)}{dt},
\end{equation}
where $g_{ij}=-\partial^2S/(\partial k_i\partial k_j)$.
Identifying the derivative
\begin{equation}\label{Jj}
J_j=\frac{d(k_j-h_j)}{dt}
\end{equation}
as a thermodynamic flow, and then comparing (\ref{RateS2}) with
(\ref{RateS1}) allows one to show that the term
\begin{equation}\label{Fj}
F_j=-\sum_{i}g_{ij}(k_i-h_i)
\end{equation}
corresponds to the thermodynamic force.

Now, assuming that the probability of a fluctuation in our
ensemble is given by the Einstein formula $P(d_iS)\sim\exp[d_iS]$
one can show that elements of the matrix $\mathbf{g^{-1}}$ (which
is the inverse of $\mathbf{g}$) describe correlations between
fluctuations \cite{Mazur,Kondepundi}
\begin{equation}\label{gm1ij}
g_{ij}^{-1}=\langle (k_i-h_i)(k_j-h_j)\rangle=\langle
k_ik_j\rangle-h_ih_j.
\end{equation}
At this point it is also worth to stress that from a physical
point of view the parameters $g_{ij}^{-1}$ correspond to
generalized susceptibilities $\chi_{ij}^{(\theta)}=-\partial
h_i/\partial\theta_j$ (see Eq.~(39) in \cite{FronczakFluct}),
which measure the response of $h_i$ to the variation of the field
$\theta_j$. Having the ensemble averages \cite{info1}
\begin{equation}\label{kikj}
\langle k_ik_j\rangle=h_i\left(1-\frac{h_i\langle
h^2\rangle}{\langle h\rangle^2N}\right)\delta_{ij}+h_ih_j,
\end{equation}
one immediately finds that in sparse and uncorrelated networks
described by the Hamiltonian (\ref{H1}), for which $\langle
h^2\rangle/\langle h\rangle\leq\ln N$
\cite{FronczakFluct,BogunaCutOff}, the matrix $\mathbf{g}$ is
diagonal
\begin{equation}\label{gm1ij2}
g_{ij}\simeq\frac{\delta_{ij}}{h_i}.
\end{equation}

\begin{figure} \epsfxsize=8.5cm \epsfbox{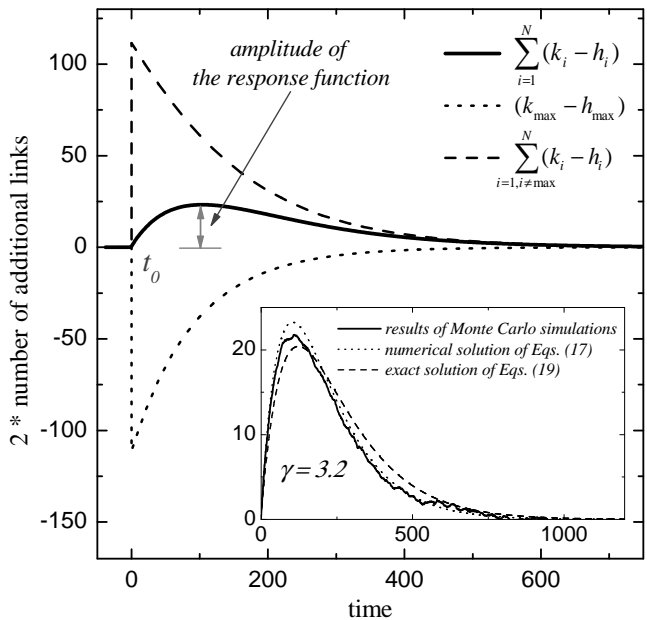}
\caption{Main stage: Schematic picture illustrating behavior of a
network after the perturbation consisting in a sudden rewiring of
all links attached to the most connected node to other nodes.
Subset: Response function of scale-free network of sizes $N=512$
with $\gamma=3.2$. (All Monte Carlo simulations presented in the
paper have been averaged $10^6$ times.)}\label{fig1}
\end{figure}

The last result is interesting for two reasons. Firstly, it allows
to simplify the expression for the thermodynamic force $F_j$
acting on the node $j$ when the studied networks are thrown out of
equilibrium. Namely, inserting (\ref{gm1ij2}) into (\ref{Fj}) one
finds that the force is equivalent to the normalized fluctuation
on the considered node
\begin{equation}\label{Fj1}
F_j=\frac{h_j-k_j}{h_j}.
\end{equation}
Secondly, it shows that correlations between fluctuations on
various nodes are negligibly small. Although at first glance the
remark seems to contradict the expected cross effects, further in
the paper we show that the effects consisting in cascading
development of different flows between the nodes do really exist
in the considered networks.

In the following, in order to examine the mentioned cross effects
we will write the rate equation for $k_j-h_j$, which will make
possible the detailed analysis of the thermodynamic flows $J_j$
(\ref{Jj}) in the considered ensemble. Before, however, we proceed
with this equation let us discuss structural and dynamical
properties of the studied networks. First, since the networks are
uncorrelated the probability of a link between any pair of nodes
$i$ and $j$ with degrees respectively equal to $k_i$ and $k_j$ is
given by $p_{ij}=k_ik_j/(\langle k\rangle N)$. Next, due to the
fact that the networks are close to equilibrium one can assume
that their dynamics after a small perturbation is the same as
their dynamics in equilibrium. One can expect that the analyzed
networks make only small steps in the configuration space
$\mathcal{G}$ forming a sort of a reasonable physical trajectory,
along which successive networks $G$ appear with probabilities
proportional to their weights, that is, proportional to
$e^{-H(G)}$ (\ref{PG}). The simplest and physically the most
reasonable method providing such a sampling is known as Metropolis
algorithm \cite{bookNewman}. In the algorithm the ratio
\begin{equation}\label{w}
w=\frac{P(G_1)}{P(G_2)}=\frac{e^{-H(G_1)}}{e^{-H(G_2)}}=e^{-\Delta
H}
\end{equation}
is interpreted as the probability of making a transition from one
network configuration $G_1$ to the other configuration $G_2$ (if
$\Delta H<0$ then $w>1$ and such a transition is always accepted).
The considered difference between the two configurations $G_1$ and
$G_2$ should not be too large, since then the acceptance
probability $w$ would be small.

Now, having in mind the expounded properties of the considered
ensemble, and assuming that during a single time step only one
link may be added or removed from the network one can easily write
the rate equation for $k_j-h_j$
\begin{eqnarray}\nonumber
\frac{\partial(k_j-h_j)}{\partial
t}=\frac{1}{\binom{N}{2}}\sum_{i\neq
j}\left[(-1)\frac{k_ik_j}{\langle k\rangle
N}\right.\min[e^{\theta_i+\theta_j},1]+
\\(+1)\left.\left(1-\frac{k_ik_j}{\langle k\rangle
N}\right)\min[e^{-(\theta_i+\theta_j)},1]\right].\label{eqA}
\end{eqnarray}
The first term on the right-hand side of Eq.~(\ref{eqA})
corresponds to node's degree decrement by a link removal, and
respectively the second term represents node's degree increment by
a link addition. At the moment, our aim is to reformulate the last
equation into the form similar to relation (\ref{FlowGen}). In
order to do it let us recall two properties of the analyzed
ensemble (\ref{H1}), which have been proved in
\cite{FronczakFluct}. The first property $\langle k\rangle=\langle
h\rangle$ is trivial and does not require any comment. The second
property, that is of our interest, relates the expected node's
degree $h_j$ with its conjugated field $\theta_j$, i.e. $h_j\simeq
e^{-\theta_j}\sqrt{\langle h\rangle N}$. The last expression is
only true in sparse and uncorrelated networks for which fields
$\{\theta_i\}$ conjugated to nodes' degrees are positive. Putting
the mentioned expressions into (\ref{eqA}), after some algebra one
gets a new rate equation
\begin{eqnarray}\nonumber
\frac{\partial(k_j-h_j)}{\partial
t}=&-&\frac{2}{N^2}\left[k_j\left(1+\frac{h_j\langle
h^2\rangle}{N\langle h\rangle^2}\right)-h_j\right]\\\label{eqB}&
-&\frac{2k_jh_j}{\langle h\rangle^2N^4}\sum_{i\neq j}h_i(k_i-h_i),
\end{eqnarray}
which after putting $k_j=h_j$ in the second term (since we operate
in the vicinity of equilibrium the assumption is reasonable)
simplifies to the desired form (\ref{FlowGen})
\begin{equation}\label{eqC}
\frac{\partial(k_j-h_j)}{\partial
t}=\frac{2h_j}{N^2}\left(\frac{h_j-k_j}{h_j}\right)+\sum_{i\neq
j}\frac{2h_i^2h_j^2}{\langle h\rangle^2N^4}
\left(\frac{h_i-k_i}{h_i}\right),
\end{equation}
having the exact solution
\begin{equation}\label{eqCsol}
\vec{k}(t)-\vec{h}=e^{-\mathbf{L}\mathbf{g}t}\left(\vec{k}(t_0)-\vec{h}\right),
\end{equation}
and providing us with the matrix of phenomenological coefficients
$\mathbf{L}$ describing non-equilibrium phenomena occurring in the
considered networks
\begin{displaymath}
L_{ij}=\left\{
\begin{array}{lll}
\dfrac{2h_i}{N^2}&\mbox{      for      }&i=j\\
\dfrac{2h_i^2h_j^2}{\langle h\rangle^2N^4}&\mbox{      for
    }&i\neq j
\end{array} \right..
\end{displaymath}

\begin{figure} \epsfxsize=7.5cm \epsfbox{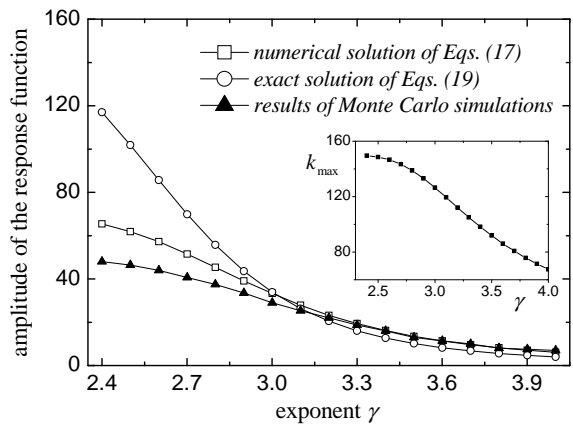}
\caption{Amplitude of the response function versus $\gamma$ in
scale-free networks of size $N=512$.}\label{fig3}
\end{figure}

Now, let us discuss results of the last paragraph. At the
beginning let us note that the equation (\ref{eqC}) clearly shows
that cross effects do really exist in complex networks.
Furthermore, the obtained matrix $\mathbf{L}$ is symmetrical. It
means that the Onsager relations (\ref{OnsagerRel}) hold in the
studied networks, i.e. the effect of a normalized fluctuation
occurring in one node $F_i$ (\ref{Fj1}) on the flow which is
induced in another node $J_j$ (\ref{Jj}) is the same as the effect
of $F_j$ on $J_i$, regardless of the nodes' degrees $h_i$ and
$h_j$. Note also that the equation (\ref{eqC}) can be written as
follows
\begin{equation}
J_j=J_j^{(j)}+\sum_{i\neq j}J_j^{(i)},
\end{equation}
revealing the multi-component nature of the analyzed flows. The
partial flows introduced in the last expression can be easily
identified from the initial equation (\ref{eqC}). They
respectively stand for flows $J_j^{(i)}=L_{ji}F_i$ generated on
the node $j$ by other nodes $i\neq j$, and for the flow
$J^{(j)}_j=L_{jj}F_j$ induced on the node by itself. A simple
comparison of the flows shows that in the studied case of sparse
and uncorrelated networks (\ref{H1}) the following relation holds
\begin{equation}\label{Jjrel}
\forall_{i\neq j} J_j^{(j)}\gg J_j^{(i)},
\end{equation}
which stems from the analogous relation between Onsager
coefficients, i.e. $\forall_{i\neq j}\;L_{jj}\gg L_{ij}$. The
above relation causes that the partial flows $J_j^{(i)}$, giving
rise to cross effects, are much smaller than the local flow
$J_j^{(j)}$. In fact, the only networks for which the total effect
of the cross flows is considerable are scale-free networks, in
which highly connected nodes appear.

Therefore, to numerically verify the obtained results we have
analyzed behavior of scale-free networks (i.e. networks
characterized by a power law distribution of the desired nodes'
degrees $P(h)\sim h^{-\gamma}$, which $2.4\leq\gamma\leq 4$) after
a sudden rewiring of all links attached to the node with the
highest degree $k_{max}$ to other nodes. Schematic illustration of
the network response to this externally applied disturbance is
shown in Fig.~\ref{fig1}. The cross effects manifest themselves in
a number of additional links which appear in the network during
its return to equilibrium. In order to quantify the effects and
check the correctness of our calculations we have measured the
amplitude of the response function (see Fig.~\ref{fig1}) obtained
from Monte Carlo simulations and compare it with both numerical
solution of the set of initial rate Eqs.~(\ref{eqA}) and the exact
solution (\ref{eqCsol}) of the set of simplified Eqs.~(\ref{eqC})
(see subset in Fig.~\ref{fig1}). The results are presented in
Fig.~\ref{fig3}. One can see that for $\gamma\geq 3$ our
analytical calculations fit numerical results very well. The
visible discrepancy between the numerical results and their
theoretical predictions for $\gamma<3$ is due to the fact that the
applied formalism does not take into account degree correlations
which spontaneously develop in scale-free networks with $\gamma<3$
(see comment after Eq.~(28) in \cite{FronczakFluct}).

In summary, in this paper we present Onsager formalism applied to
random networks with arbitrary degree distribution. Using the
well-known methods of non-equilibrium thermodynamics we identify
thermodynamic forces and their conjugated flows induced in
networks as a result of single node degree perturbation. The
forces and the flows can be understood as a response of the system
to events, such as random removal of nodes or intentional attacks
on them. We show that cross effects (such as thermodiffusion, or
thermoelectric phenomena), in which one force may not only give
rise to its own corresponding flow, but to many other flows, can
be observed also in complex networks.

Finally, since the science of complex networks is a genuinely
multidisciplinary domain, the approach if applied to social,
economic, or even biological networks may open new horizons for
the sciences, as it would provide them with a completely new
understanding of how rumors, information, marketing, or crises can
spread through these systems causing small, medium or large
responses. Moreover, if one can identify social (economic)
equivalents of thermodynamic forces and flows, a social (economic)
analogue of thermodynamic cross effects, underlying complexity of
the socio-economic systems, will be within the grasp. We hope that
the approach introduced in the paper will serve as a practical
starting point for exploring a variety of non-equilibrium
network-driven phenomena.

The work was funded in part by the State Committee for Scientific
Research in Poland under Grant 1P03B04727 (A.F.), the European
Commission Project CREEN FP6-2003-NEST-Path-012864 (P.F.), and by
the Ministry of Education and Science in Poland under Grant
134/E-365/6.PR~UE/DIE 239/2005-2007 (J.A.H.).


\begin{thebibliography}{6}
\bibitem{OnsagerPR} L. Onsager, Phys. Rev. {\bf 37}, 405 (1931); Phys. Rev {\bf 38}, 2265 (1931).
\bibitem{Mazur} S.R. Groot and P. Mazur, {\it Non-equilibrium thermodynamics}, North Holland, Amsterdam (1982).
\bibitem{Kondepundi} D. Kondepundi and I. Priggine, {\it Modern thermodynamics}, John Wiley $\&$ Sons, New York (1998).
\bibitem{net1} S. Bornholdt and H.G. Schuster, {\it Handbook of graphs and networks}, Wiley-Vch (2002).
\bibitem{net2} S.N. Dorogovtsev and J.F.F. Mendes, {\it Evolution of networks}, Oxford Univ. Press (2003).
\bibitem{pmodels1} D. Strauss, SIAM Rev. {\bf 28}, 513 (1986).
\bibitem{pmodels2} O. Frank and D. Strauss, J. Am. Stat. Assoc. {\bf 81}, 832 (1986).
\bibitem{Park04} J. Park and M.E.J. Newman, Phys. Rev. E {\bf 70}, 066117 (2004).
\bibitem{FronczakFluct} A. Fronczak, P. Fronczak and J.A. Ho\l yst, Phys. Rev. E {\bf 73}, 016108 (2006)
\bibitem{Garla06} D. Garlaschelli and M.I. Loffredo, Phys. Rev. E {\bf 73}, 015101(R) (2006).
\bibitem{Jaynes57} E.T. Jaynes, Phys. Rev. {\bf 106}, 620 (1957); Phys. Rev. {\bf 108}, 171 (1957).
\bibitem{New01} M.E.J. Newman, S.H. Strogatz, and D.J. Watts, Phys. Rev. E {\bf 64}, 026118 (2001).
\bibitem{Pastor01} R. Pastor-Satorras, and A. Vespignani, Phys. Rev. Lett. {\bf 86}, 3200 (2001).
\bibitem{HavlinPRLe} R. Cohen  et al., Phys. Rev. Lett. {\bf 85}, 4626 (2000); Phys. Rev. Lett. {\bf 86}, 3682 (2001).
\bibitem{BANatur} R. Albert, H. Jeong, and A.L. Barab\'{a}si, Nature {\bf 406}, 378 (2000).
\bibitem{MotterPRL} A.E. Motter, Phys. Rev. Lett. {\bf 93}, 098701 (2004).
\bibitem{BianconiPRE} G. Bianconi, and M. Marsili, Phys. Rev. E {\bf 70}, 035105(R) (2004).
\bibitem{FronczakSOC} P. Fronczak, A. Fronczak, and J.A. Ho\l yst, Phys. Rev. E {\bf 74}, 026121 (2006).
\bibitem{info1} $\langle k_ik_j\rangle=\sum_{x,y}\langle
a_{ix}a_{jy}\rangle$, where $a_{ix}$ and $a_{iy}$ represent
entries of the adjacency matrix, and $\langle
a_{ix}a_{jy}\rangle=(\langle a_{ix}\rangle-\langle
a_{ix}\rangle^2)\delta_{ij}\delta_{xy}+\langle
a_{ix}\rangle\langle a_{jy}\rangle$ Eq.~(7) in \cite{Dorog03},
whereas $\langle a_{ij}\rangle=h_ih_j(\langle h\rangle N)^{-1}$
Eq.~(26) in \cite{FronczakFluct}.
\bibitem{Dorog03} S.N. Dorogovtsev, cond-mat/0308336.
\bibitem{BogunaCutOff} M. Bogu\~{n}\'{a}, R. Pastor-Satorras, and A. Vespignani, Eur. Phys. J. B {\bf 38}, 205 (2004).
\bibitem{bookNewman} M.E.J. Newman and G.T. Barkema, {\it Monte Carlo methods in statistical physics}, Oxford Univ. Press (2006).

\end{thebibliography}
\end{document}